\documentclass[conference,a4paper]{IEEEtran}
\usepackage{amsfonts}
\usepackage{amstext}
\usepackage{amssymb}
\usepackage{amsmath}
\usepackage{graphicx}
\usepackage{enumerate}

\newtheorem{remark}{Remark}
\newtheorem{theorem}{Theorem}

\DeclareMathOperator*{\argmin}{arg\,min}

\newcommand{\nodeA}{A}
\newcommand{\nodeB}{B}
\newcommand{\relay}{R}

\newcommand{\X}{X}
\newcommand{\Y}{Y}
\newcommand{\Z}{Z}
\newcommand{\x}{x}
\newcommand{\y}{y}
\newcommand{\z}{z}
\newcommand{\alphx}{\mathcal{\X}}

\newcommand{\hatX}{\hat{\X}}
\newcommand{\hatY}{\hat{\Y}}

\newcommand{\D}{D}
\newcommand{\fd}{d}

\newcommand{\msgM}{M}
\newcommand{\msgm}{m}

\newcommand{\rate}{R}
\newcommand{\rdreg}{\mathfrak{\rate}}
\newcommand{\ratereg}{\mathcal{\rate}}

\newcommand{\auxU}{U}
\newcommand{\auxV}{V}
\newcommand{\auxW}{W}
\newcommand{\auxS}{S}
\newcommand{\auxu}{u}
\newcommand{\auxv}{v}
\newcommand{\auxw}{w}
\newcommand{\auxs}{s}

\newcommand{\indxi}{i}
\newcommand{\indxt}{j}
\newcommand{\indxtt}{k}

\newcommand{\encod}{\phi}
\newcommand{\decod}{\Psi}

\begin{document}

\sloppy

\title{Interactive Relay Assisted Source Coding}

\author{
   \IEEEauthorblockN{Farideh Ebrahim Rezagah and Elza Erkip}
   \IEEEauthorblockA{Dept. of ECE, Polytechnic Institute of NYU\\
     Email: fer216@nyu.edu, elza@poly.edu}
 }

\maketitle

\begin{abstract}
   This paper investigates a source coding problem in which two terminals communicating through a relay wish to estimate one another's source within some distortion constraint. The relay has access to side information that is correlated with the sources. Two different schemes based on the order of communication, \emph{distributed source coding/delivery} and \emph{two cascaded rounds}, are proposed and inner and outer bounds for the resulting rate-distortion regions are provided. Examples are provided to show that neither rate-distortion region includes the other one.

\end{abstract}

\section{Introduction}

\begin{figure}[htbp]
   \centering
   \includegraphics[trim=10cm 11.5cm 10cm 2.3cm, clip, width=0.4\textwidth]{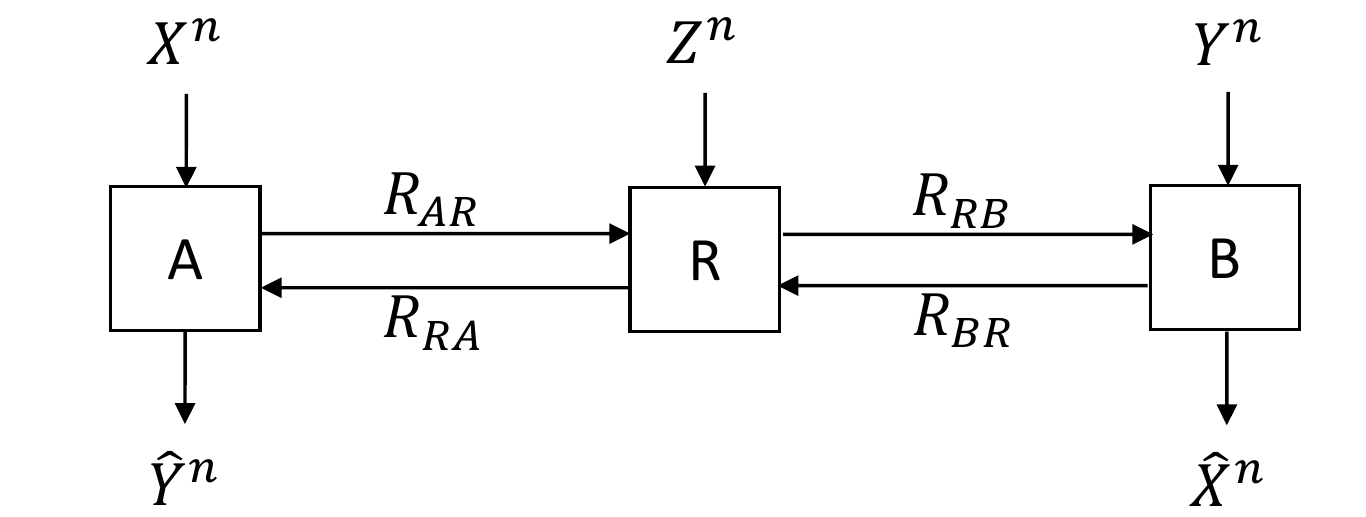}
   \caption{Interactive relay assisted source coding.}
   \label{fig:system model z}
 \end{figure}

Consider a distributed source coding problem in which two terminals A and B, each having access to a correlated source sequence $\X^n$ and $\Y^n$ respectively, wish to obtain lossless or lossy estimates of each other's sources. The terminals are only able to interact over rate limited directed links through a relay ($\relay$), who observes a correlated sequence $\Z^n$. This scenario, which we call \textit{interactive relay assisted source coding} is illustrated in Fig \ref{fig:system model z}. While in general multiple communication rounds through the relay are possible as in interactive source coding with multiple rounds \cite{Kaspi1985}; here we only assume that each link is used once.

One can envision various communication schemes depending on the order of communication. For example, first $\nodeA$ and $\nodeB$ can communicate with $\relay$, which then processes both signals and sends information back to the terminals. This scheme breaks the problem of interactive relay assisted source coding into a distributed source coding (DSC) phase \cite{NIT} with no reconstruction constraints at the relay, and a delivery (broadcast) phase \cite{Kimura2008}. We will call this scheme \emph{DSC/delivery}. Note that while this is similar to the scenario considered in \cite{SuElGamal2010}, a main difference is that in the delivery phase, the relay has two separate rate-constrained links to the terminals. Another strategy is for $\nodeA$ to first forward information to $\nodeB$ through $\relay$, which then obtains its estimate $\hatX^n$ and forwards information back to $\nodeA$ through $\relay$. This scheme puts together two cascaded communication rounds \cite{Yamamoto1981} - one in each direction -  and will be called \emph{two cascaded rounds}. Our goal in this paper is to study inner and outer bounds for the rate-distortion regions of above two strategies and examine some conditions under which these bounds are tight. We also argue that neither strategy strictly dominates the other in general.

Interactive relay assisted source coding is related to several other problems investigated in the literature. The first step in DSC/delivery is closely related to lossy DSC, which is only solved for certain cases \cite{Wagner2008,Courtade2012}. The complementary delivery problem, which constitutes the broadcast phase of DSC/delivery scheme, is investigated in \cite{Kimura2008}.

The cascade source coding problem, which considers only one round of the two cascaded round scheme (from $\nodeA$ to $\relay$ and then to $\nodeB$) is introduced by Yamamoto \cite{Yamamoto1981}. In \cite{Yamamoto1981} there is no side information, and $\relay$ and $\nodeB$ are interested in different lossy reconstructions of the source. Vasudevan et al. \cite{VasudevanTianDiggavi2006} further investigated this problem by letting $\relay$ and $\nodeB$ to have access to side information $\Z$ and $\Y$ respectively, where $\X-\Y-\Z$. They considered the cases where $\nodeA$ may or may not have access to $\Z$ and $\Y$. In \cite{CuffSuElGamal2009} $\nodeB$ has side information and there are no distortion constraints at $\relay$.

This paper is organized as follows. In Section \ref{setup} we present the system model, notation and the two schemes, DSC/delivery and two cascaded rounds. For each scheme we characterize inner and outer bounds in Sections \ref{dscd} and \ref{cascade}. In Section \ref{discussion} we introduce some examples to show that although these bounds are not tight for general sources, they can be used to prove that neither scheme is in general dominant.



\section{Problem Set-Up}\label{setup}

\subsection{System Model}

We consider interactive relay assisted source coding shown in Fig \ref{fig:system model z}. Terminal $\nodeA$ observes source $\X^n$, terminal $\nodeB$ observes $\Y^n$, and the relay, $\relay$, observes $\Z^n$. The sequences are drawn $i.i.d.\sim p(\x,\y,\z)$. The objective of terminal $\indxt$, $\indxt=\nodeA,\nodeB$ is to recover the source at terminal $\indxtt$, $\indxtt=\nodeB,\nodeA$, $\indxt\not=\indxtt$ in a lossy fashion. To achieve this, messages $\msgM_{\nodeA\relay}$, $\msgM_{\relay\nodeB}$, $\msgM_{\nodeB\relay}$, and $\msgM_{\relay\nodeA}$ at rates $\rate_{\nodeA\relay}$, $\rate_{\relay\nodeB}$, $\rate_{\nodeB\relay}$, and $\rate_{\relay\nodeA}$ respectively are communicated among the terminals, where $\msgM_{\indxt\indxtt}$ is the message sent from $\indxt$ to $\indxtt$. We consider the following two schemes, each with a specified communication order:
\begin{enumerate}[(a)]

  \item \textit{DSC/delivery}

  In this scheme, the messages are sent in the following order: First $\nodeA$ and $\nodeB$ generate and send $\msgM_{\nodeA\relay}$ and $\msgM_{\nodeB\relay}$ respectively, to $\relay$. This is the DSC stage. The order of these two messages does not impact the resulting distortions at the terminals. After $\relay$ gets both messages, it generates $\msgM_{\relay\nodeA}$ and $\msgM_{\relay\nodeB}$ and sends them to $\nodeA$ and $\nodeB$ respectively. Similar to the DSC stage, these two massages can be sent in any order. More formally the encoding functions are $\msgM_{\nodeA\relay}=\encod^{(n)}_{\nodeA\relay}(\X^n)$, $\msgM_{\nodeB\relay}=\encod^{(n)}_{\nodeB\relay}(\Y^n)$, $\msgM_{\relay\nodeA}=\encod^{(n)}_{\relay\nodeA}(\Z^n,\msgM_{\nodeA\relay},\msgM_{\nodeB\relay})$ and $\msgM_{\relay\nodeB}=\encod^{(n)}_{\relay\nodeB}(\Z^n,\msgM_{\nodeA\relay},\msgM_{\nodeB\relay})$.\\

 \item \textit{Two cascaded rounds}

The order in which the messages are sent is given by $\nodeA$ to $\relay$, $\relay$ to $\nodeB$, $\nodeB$ to $\relay$ and then finally $\relay$ to $\nodeA$. Similarly, the communication can start from $\nodeB$, but in the rest of the paper we will assume $\nodeA$ starts the communication. This can be mathematically described by the following encoding functions: $\msgM_{\nodeA\relay}=\encod^{(n)}_{\nodeA\relay}(\X^n)$, $\msgM_{\relay\nodeB}=\encod^{(n)}_{\relay\nodeB}(\Z^n,\msgM_{\nodeA\relay})$, $\msgM_{\nodeB\relay}=\encod^{(n)}_{\nodeB\relay}(\Y^n,\msgM_{\relay\nodeB})$ and $\msgM_{\relay\nodeA}=\encod^{(n)}_{\relay\nodeA}(\Z^n,\msgM_{\nodeA\relay},\msgM_{\nodeB\relay})$.

\end{enumerate}

The decoding functions in both schemes are given by: $\hatY^n=\decod^{(n)}_\nodeA(\X^n,\msgM_{\relay\nodeA})$ and $\hatX^n=\decod^{(n)}_\nodeB(\Y^n,\msgM_{\relay\nodeB})$. Also note that in both schemes, $\msgM_{\indxt\indxtt}\in\{1,2,\dots,2^{n\rate_{\indxt\indxtt}}\}$ where $\rate_{\indxt\indxtt}$ is the directed communication rate between terminals $\indxt$ and $\indxtt$.

For any of the schemes the tuple $(\encod^{(n)}_{\nodeA\relay},\encod^{(n)}_{\nodeB\relay},\encod^{(n)}_{\relay\nodeA},\encod^{(n)}_{\relay\nodeB},$ $\decod^{(n)}_\nodeA,\decod^{(n)}_\nodeB)$ is called an $(n,\rate_{\nodeA\relay},\rate_{\nodeB\relay},\rate_{\relay\nodeA},\rate_{\relay\nodeB},\D_\nodeA,\D_\nodeB)$-\textit{code} if the produced sequences $\hatX^n$ and $\hatY^n$ satisfy $\frac{1}{n}\sum_{\indxi=1}^n E\fd_\nodeA(\Y_\indxi,\hatY_\indxi)\leq\D_\nodeA$ and $\frac{1}{n}\sum_{\indxi=1}^n E\fd_\nodeB(\X_\indxi,\hatX_\indxi)\leq\D_\nodeB$, where $\fd_\indxt(.,.)$, $\indxt =\nodeA,\nodeB$ are single letter distortion functions at the corresponding terminals. The rate and distortion tuple $(\rate_{\nodeA\relay},\rate_{\nodeB\relay},\rate_{\relay\nodeA},\rate_{\relay\nodeB},\D_\nodeA,\D_\nodeB)$ is \textit{achievable} if for any $\epsilon\geq 0$ and $n$ sufficiently large there exists an $(n,\rate_{\nodeA\relay}+\epsilon,\rate_{\nodeB\relay}+\epsilon,\rate_{\relay\nodeA}+\epsilon,\rate_{\relay\nodeB}+\epsilon,\D_\nodeA+\epsilon,\D_\nodeB+\epsilon)$-code.

The set of all achievable $(\rate_{\nodeA\relay},\rate_{\nodeB\relay},\rate_{\relay\nodeA},\rate_{\relay\nodeB},\D_\nodeA,\D_\nodeB)$ is denoted by $\rdreg_{sch}$, where $sch$ refers to one of the two schemes above. The \textit{interactive relay assisted rate-distortion region} for that scheme is given by:
\begin{IEEEeqnarray*}{rl}
  \ratereg_{sch}(\D_\nodeA,\D_\nodeB)&\triangleq \bigg\{(\rate_{\nodeA\relay},\rate_{\nodeB\relay},\rate_{\relay\nodeA},\rate_{\relay\nodeB})|\\
  &(\rate_{\nodeA\relay},\rate_{\nodeB\relay},\rate_{\relay\nodeA},\rate_{\relay\nodeB},\D_\nodeA,\D_\nodeB)\in\rdreg_{sch}\bigg\}
\end{IEEEeqnarray*}

\begin{figure}[htbp]
   \centering
   \includegraphics[trim=10.5cm 13cm 13cm 1cm, clip, width=0.3\textwidth]{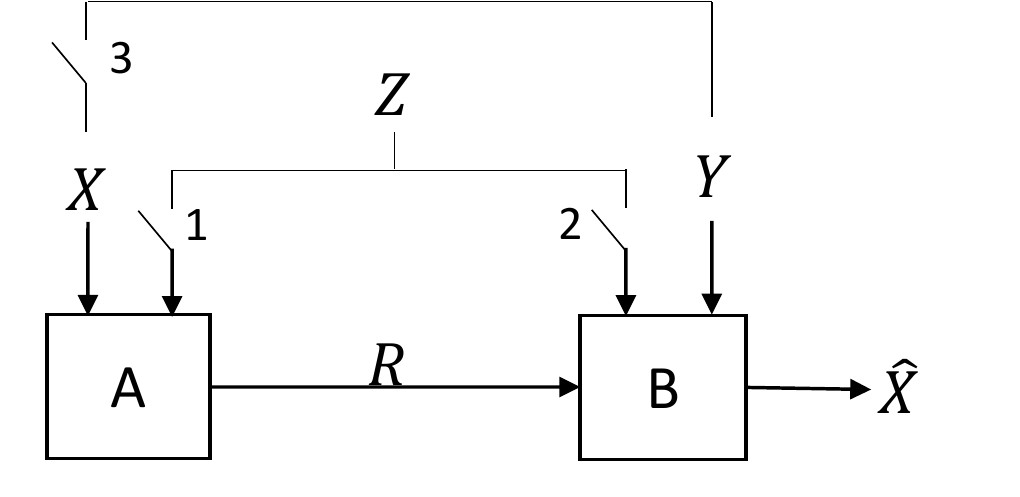}
   \caption{Two terminal source coding with encoder and side information.}
   \label{fig:modifiedWZ}
 \end{figure}

\subsection{Notation}

  We will use the following notation based on the source coding set-up in Fig \ref{fig:modifiedWZ}. When all switches are open, this set-up becomes identical to Wyner-Ziv problem \cite{WynerZiv1976}, with rate-distortion function $\rate_{\X|\Y}^{WZ}(\D)$. We let $\rate_{\X;\Z|\Y}^{WZ^*}(\D)$ be the rate-distortion function when switch 1 is closed, while switches 2 and 3 are still open. We also define $\rate_{\X;\Z|\Y\Z}^{WZ^*}(\D)$ to be the rate-distortion function when switches 1 and 2 are closed while 3 is still open. Finally when all switches are closed we get $\rate_{\X|\Y\Z}(\D)$, the rate-distortion function with side information $\Y\Z$ available both at the encoder and the decoder. With these definitions we have:
\begin{IEEEeqnarray*}{l}
    \rate_{\X|\Y}^{WZ}(\D)\geq\rate_{\X;\Z|\Y}^{WZ^*}(\D)\geq\rate_{\X;\Z|\Y\Z}^{WZ^*}(\D)\geq\rate_{\X|\Y\Z}(\D)
\end{IEEEeqnarray*}

\section{DSC/Delivery}\label{dscd}

\subsection{Outer Bound}

\begin{theorem}\label{DSC/deliveryOB}
   If $(\rate_{\nodeA\relay},\rate_{\nodeB\relay},\rate_{\relay\nodeA},\rate_{\relay\nodeB})$ is in the rate-distortion region of the DSC/delivery scheme, $\ratereg_{DSCD}(\D_\nodeA,\D_\nodeB)$, then
  \begin{IEEEeqnarray*}{l}
    \rate_{\nodeA\relay}\geq I(\X;\auxU_1|\Z,\Y)\label{DSC-AR}\\
    \rate_{\nodeB\relay}\geq I(\Y;\auxU_2|\Z,\X)\label{DSC-BR}\\
    \rate_{\relay\nodeA}\geq I(\Y;\auxV_2|\X,\auxU_1)\label{DSC-RA}\\
    \rate_{\relay\nodeB}\geq I(\X;\auxV_1|\Y,\auxU_2)\label{DSC-RB}
  \end{IEEEeqnarray*}
  for some $(\auxU_1,\auxU_2,\auxV_1,\auxV_2)$ where $\auxU_1-\X-(\Y,\Z)$, $\auxU_2-\Y-(\X,\Z)$, and $(\auxV_1,\auxV_2)-(\Z,\auxU_1,\auxU_2)-(\X,\Y)$ and decoding functions $\hatY=\decod_\nodeA(\X,\auxU_1,\auxV_2)$ and $\hatY=\decod_\nodeB(\Y,\auxU_2,\auxV_1)$ such that $E\fd_\nodeA(\Y,\hatY)\leq\D_\nodeA$ and $E\fd_\nodeB(\X,\hatX)\leq\D_\nodeB$.
\end{theorem}

\begin{IEEEproof}
  We first lower bound $\rate_{\nodeA\relay}$ as follows:
  \begin{IEEEeqnarray*}{rl}
     n\rate_{\nodeA\relay}&\geq H(\msgM_{\nodeA\relay}|\Z^n,\Y^n)=I(\X^n;\msgM_{\nodeA\relay}|\Z^n,\Y^n)\\
     	&=H(\X^n|\Z^n,\Y^n)-H(\X^n|\msgM_{\nodeA\relay},\Z^n,\Y^n)\\
	 &=\sum_{\indxi=1}^{n}\left[H(\X_\indxi|\Z_\indxi,\Y_\indxi)-H(\X_\indxi|\msgM_{\nodeA\relay},\Z^n,\Y^n,\X^{\indxi-1})\right]\\
	&\geq\sum_{\indxi=1}^{n}I(\X_\indxi;\auxU_{1,\indxi}|\Z_\indxi,\Y_\indxi)
  \end{IEEEeqnarray*}
  where $\auxU_{1,\indxi}=(\msgM_{\nodeA\relay},\X^{\indxi-1},\Y_{\indxi+1}^n)$. Similarly $n\rate_{\nodeB\relay}\geq\sum_{\indxi=1}^{n}I(\Y_\indxi;\auxU_{2,\indxi}|\Z_\indxi,\X_\indxi)$ with $\auxU_{2,\indxi}=(\msgM_{\nodeB\relay},\X_{\indxi+1}^n,\Y^{\indxi-1})$. For the delivery phase we have:
  \begin{IEEEeqnarray*}{rl}
     n\rate_{\relay\nodeA}&\geq H(\msgM_{\relay\nodeA}|\msgM_{\nodeA\relay},\X^n)\geq I(\Y^n;\msgM_{\relay\nodeA}|\msgM_{\nodeA\relay},\X^n)\\
	&=\sum_{\indxi=1}^{n}I(\Y_\indxi;\msgM_{\relay\nodeA}|\msgM_{\nodeA\relay},\X^n,\Y_{\indxi+1}^n)\\
	 &=\sum_{\indxi=1}^{n}I(\Y_\indxi;\msgM_{\relay\nodeA},\X_{\indxi+1}^n|\msgM_{\nodeA\relay},\X^\indxi,\Y_{\indxi+1}^n)\\
	&\geq\sum_{\indxi=1}^{n}I(\Y_\indxi;\auxV_{2,\indxi}|\X_\indxi,\auxU_{1,\indxi})
  \end{IEEEeqnarray*}
  where $\auxV_{2,\indxi}=\msgM_{\relay\nodeA}$. Similarly with $\auxV_{1,\indxi}=\msgM_{\relay\nodeB}$, we have $n\rate_{\relay\nodeB}\geq\sum_{\indxi=1}^{n}I(\X_\indxi;\auxV_{1}|\Y_\indxi,\auxU_{2,\indxi})$. It is easy to verify that $\auxU_{1,\indxi}$, $\auxU_{2,\indxi}$, $\auxV_{1,\indxi}$ and $\auxV_{2,\indxi}$ satisfy the required Markov conditions.

Define $\decod_{\nodeA,\indxi}^{(n)}$ to be the function that maps $(\X^n,\msgM_{\relay\nodeA})$ to the $\indxi^{th}$ symbol of $\decod_\nodeA^{(n)}(\X^n,\msgM_{\relay\nodeA})$ for the code at hand. Similarly let $\decod_{\nodeB,\indxi}^{(n)}$ be the function that maps $(\Y^n,\msgM_{\relay\nodeB})$ to the $\indxi^{th}$ symbol of $\decod_\nodeB^{(n)}(\Y^n,\msgM_{\relay\nodeB})$. We next argue that there exist deterministic functions $\decod_{\nodeA,\indxi}(\X_\indxi,\auxU_{1,\indxi},\auxV_{2,\indxi})$ and $\decod_{\nodeB,\indxi}(\Y_\indxi,\auxU_{2,\indxi},\auxV_{1,\indxi})$ that achieve distortions no larger than what $\decod_{\nodeA,\indxi}^{(n)}(\X^n,\msgM_{\relay\nodeA})$ and $\decod_{\nodeB,\indxi}^{(n)}(\Y^n,\msgM_{\relay\nodeB})$ can achieve respectively. We define $\D_{\nodeA,\indxi}$ as
  \begin{IEEEeqnarray*}{rCl}
     \D_{\nodeA,\indxi}&\triangleq E&\left[\fd_\nodeA(\Y_\indxi,\hatY_\indxi)\right]\\
	&=& E_{\X^n,\Y_{\indxi}^n,\msgM_{\nodeA\relay},\msgM_{\relay\nodeA}}\left[\fd_\nodeA(\Y_\indxi,\decod_{\nodeA,\indxi}^{(n)}(\X^n,\msgM_{\relay\nodeA}))\right]\\
	 &=&E_{\X^n,\Y_{\indxi+1}^n,\msgM_{\nodeA\relay},\msgM_{\relay\nodeA}}E_{\Y_\indxi|\X^n,\Y_{\indxi+1}^n,\msgM_{\nodeA\relay},\msgM_{\relay\nodeA}}\\
	&&\left[\fd_\nodeA(\Y_\indxi,\decod_{\nodeA,\indxi}^{(n)}(\X^n,\msgM_{\relay\nodeA}))\right]
  \end{IEEEeqnarray*}
Let
  \begin{IEEEeqnarray*}{c}
    \tilde{\x}_{\indxi+1}^n(\x^\indxi,\y_{\indxi+1}^n,\msgm_{\nodeA\relay},\msgm_{\relay\nodeA})\triangleq \argmin_{\x_{\indxi+1}^n\in\alphx^{n-\indxi}}\\ \quad \quad
	 \big(E_{\Y_\indxi|\X^\indxi=\x^\indxi,\X_{\indxi+1}^n=\x_{\indxi+1}^n,\Y_{\indxi+1}^n=\y_{\indxi+1}^n,\msgM_{\nodeA\relay}=\msgm_{\nodeA\relay},\msgM_{\relay\nodeA}=\msgm_{\relay\nodeA}}\\ \quad \quad
	\left[\fd_\nodeA(\Y_\indxi,\decod_{\nodeA,\indxi}^{(n)}(\X^n,\msgM_{\relay\nodeA}))\right]\big)
  \end{IEEEeqnarray*}
and define
  \begin{IEEEeqnarray*}{c}
     \decod_{\nodeA,\indxi}(\x_\indxi,\auxu_{1,\indxi},\auxv_{2,\indxi})\triangleq\\
     \decod_{\nodeA,\indxi}^{(n)}(\x^\indxi,\tilde{\x}_{\indxi+1}^n(\x^\indxi,\y_{\indxi+1}^n,\msgm_{\nodeA\relay},\msgm_{\relay\nodeA}),\msgm_{\relay\nodeA})
  \end{IEEEeqnarray*}
It is easy to show that $\Y_\indxi-(\X_{\indxi},\auxU_{1,\indxi},\auxV_{2,\indxi})-\X_{\indxi+1}^n$. Using this Markov chain we have:
  \begin{IEEEeqnarray*}{c}
     E\left[\fd_\nodeA(\Y_\indxi,\decod_{\nodeA,\indxi}(\X_\indxi,\auxU_{1,\indxi},\auxV_{2,\indxi}))\right]\leq\D_{\nodeA,\indxi}
  \end{IEEEeqnarray*}
Similarly we can define $\decod_{\nodeB,\indxi}(\Y_\indxi,\auxU_{2,\indxi},\auxV_{1,\indxi})$. Using these single letter decoders the rest of the proof follows from convexity of the proposed outer bound.
\end{IEEEproof}

\begin{remark}\label{DSC delivery - simple outer}
  The following simple outer bound:
  \begin{IEEEeqnarray*}{l}
    \rate_{\nodeA\relay}\geq\rate_{\X|\Y\Z}^{WZ}(\D_\nodeB)\\
    \rate_{\nodeB\relay}\geq\rate_{\Y|\X\Z}^{WZ}(\D_\nodeA)\\
    \rate_{\relay\nodeA}\geq\rate_{\Y|\X}(\D_\nodeA)\\
    \rate_{\relay\nodeB}\geq\rate_{\X|\Y}(\D_\nodeB)
  \end{IEEEeqnarray*}
  which can be found by considering the cuts between each terminal $\indxt$ and super-node $(\indxtt,\relay)$, $\indxt,\indxtt=\nodeA,\nodeB$ and $\indxt\not=\indxtt$ is looser than the bound in Theorem \ref{DSC/deliveryOB}. This is because the cut-set bound is found by optimizing each rate separately with respect to auxiliary random variables, while in Theorem \ref{DSC/deliveryOB}, a joint optimization is considered.
\end{remark}

\subsection{Inner Bound}

\begin{theorem}
   Any rate vector $(\rate_{\nodeA\relay},\rate_{\nodeB\relay},\rate_{\relay\nodeA},\rate_{\relay\nodeB})$ satisfying
  \begin{IEEEeqnarray}{l}
    \rate_{\nodeA\relay}\geq I(\X;\auxU_1|\Z,\auxU_2,Q)\label{DSC i AR}\\
    \rate_{\nodeB\relay}\geq I(\Y;\auxU_2|\Z,\auxU_1,Q)\label{DSC i BR}\\
     \rate_{\nodeA\relay}+ \rate_{\nodeB\relay}\geq I(\X,\Y;\auxU_1,\auxU_2|\Z,Q)\label{DSC i AR BR}\\
    \rate_{\relay\nodeA}\geq I(\auxU_2,\Z;\auxV_2|\X,\auxU_1,Q)\label{DSC i RB}\\
    \rate_{\relay\nodeB}\geq I(\auxU_1,\Z;\auxV_1|\Y,\auxU_2,Q)\label{DSC i RA}
  \end{IEEEeqnarray}
  for some $p(\x,\y,\z)p(q)p(\auxu_1|\x,q)p(\auxu_2|\y,q)p(\auxv_1,\auxv_2|\auxu_1,\auxu_2,\z,$ $q)$ and some decoding functions $\hatY=\decod_\nodeA(\X,\auxV_2)$ and $\hatX=\decod_\nodeB(\Y,\auxV_1)$ such that $E\fd_\nodeA(\Y,\hatY)\leq\D_\nodeB$ and $E\fd_\nodeB(\X,\hatX)\leq\D_\nodeB$ is in the rate-distortion region of the DSC/delivery scheme, $\ratereg_{DSCD}(\D_\nodeA,\D_\nodeB)$.
\end{theorem}

\begin{IEEEproof}
  Terminals $\nodeA$ and $\nodeB$ use Berger-Tung coding \cite{NIT} to encode their sources into $\auxU_1$ and $\auxU_2$ respectively. The relay exploits its side information $\Z$ and recovers $(\auxU_1,\auxU_2)$ provided that (\ref{DSC i AR})-(\ref{DSC i AR BR}) are satisfied. The relay then uses Wyner-Ziv coding to communicate with each terminal leading to (\ref{DSC i RA}) and (\ref{DSC i RB}).
\end{IEEEproof}

\section{Two Cascaded Rounds}\label{cascade}

\subsection{Outer Bound}

\begin{theorem}\label{cascade outer bound}
   If $(\rate_{\nodeA\relay},\rate_{\relay\nodeB},\rate_{\nodeB\relay},\rate_{\relay\nodeA})$ is in the rate-distortion region of the two cascaded rounds scheme, $\ratereg_{TCR}(\D_\nodeA,\D_\nodeB)$, then
  \begin{IEEEeqnarray}{l}
    \rate_{\nodeA\relay}\geq\rate_{\X|\Y\Z}^{WZ}(\D_\nodeB)\label{r1cascade}\\
    \rate_{\relay\nodeB}\geq\rate_{\X;\Z|\Y}^{WZ^*}(\D_\nodeB)\label{r2cascade}\\
    \rate_{\nodeB\relay}\geq\rate_{\Y;\X|\Z\X}^{WZ^*}(\D_\nodeA)\label{r3cascade}\\
    \rate_{\relay\nodeA}\geq\rate_{\Y|\X}(\D_\nodeA)\label{r4cascade}
  \end{IEEEeqnarray}
\end{theorem}
\begin{IEEEproof}
   Considering the cut between $\nodeA$ and the super-node $(\relay,\nodeB)$, we get (\ref{r1cascade}). The inequality (\ref{r2cascade}) comes from considering the cut between super-source $(\nodeA,\relay)$ and the terminal $\nodeB$.

   For communication between $\nodeB$ and $\relay$, consider the cut between $\nodeB$ and the super-node $(\relay,\nodeA)$, and assume that $\X$ is known to all terminals, which leads to (\ref{r3cascade}). Now consider the cut that separates $\nodeA$ and the super-source $(\relay,\nodeB)$. With $\X$ known to all terminals, having the extra side information $\Z$ at the encoder does not help, hence we get (\ref{r4cascade}).
\end{IEEEproof}

\subsection{Inner Bound}

\begin{theorem} \label{cascade inner 1}
  Any rate vector $(\rate_{\nodeA\relay},\rate_{\relay\nodeB},\rate_{\nodeB\relay},\rate_{\relay\nodeA})$ satisfying:
  \begin{IEEEeqnarray}{rl}
    \rate_{\nodeA\relay}\geq& I(\X;\auxU_0,\auxU_1,\auxW_1|\Z)+I(\X;\auxV_1|\Y,\auxU_1)\label{R1cascade}\\
    \rate_{\relay\nodeB}\geq& I(\X;\auxU_1,\auxV_1|\Y)+I(\Z,\auxW_1;\auxS_1|\Y,\auxU_1,\auxV_1)\label{R2cascade}\\
    \rate_{\nodeB\relay}\geq& I(\Y;\auxU_2,\auxW_2|\Z,\auxU_0,\auxU_1,\auxW_1,\auxS_1)\nonumber\\
    		&+I(\Y;\auxV_2|\X,\auxU_0,\auxU_1,\auxV_1,\auxW_1,\auxU_2) \label{R3cascade}\\
    \rate_{\relay\nodeA}\geq& I(\Y;\auxU_2,\auxV_2|\X,\auxU_0,\auxU_1,\auxV_1,\auxW_1)\nonumber\\
    		&+I(\Z,\auxW_2;\auxS_2|\X,\auxU_0,\auxU_1,\auxV_1,\auxW_1,\auxU_2,\auxV_2) \label{R4cascade}
  \end{IEEEeqnarray}
  for some $p(\x,\y,\z)\linebreak[1]p(\auxu_0,\auxu_1,\auxv_1,\auxw_1|\x)\linebreak[1]p(\auxs_1|\z,\auxu_0,\auxu_1,\auxw_1)\linebreak[1]p(\auxu_2,\linebreak[1]\auxv_2,\linebreak[1]\auxw_2|\y,\auxu_1,\auxv_1,\auxs_1)\linebreak[1]p(\auxs_2|\z,\auxu_0,\auxu_1,\auxw_1,\auxs_1,\auxu_2,\auxw_2)$ and some decoding functions $\hatX=\decod_\nodeB(\Y,\auxU_1,\auxV_1,\auxS_1)$ and $\hatY=\decod_\nodeA(\X,\auxU_0,\auxU_1,\auxV_1,\auxW_1,\auxU_2,\auxV_2,\auxS_2)$ such that $E\fd_\nodeA(\Y,\hatY)\leq\D_\nodeA$ and $E\fd_\nodeB(\X,\hatX)\leq\D_\nodeB$ is in the rate-distortion region of two cascaded rounds scheme, $\ratereg_{TCR}(\D_\nodeA,\D_\nodeB)$.
\end{theorem}

\begin{IEEEproof}
  From $\nodeA$ to $\relay$ to $\nodeB$ we consider a communication scheme consisting of four separate flows: \textit{private message for $\relay$}, \textit{simple forward}, \textit{recover and forward} and \textit{recompression at $\relay$}. As the name suggests private message is the part of information that is only intended to be received by $\relay$. Simple forward refers to $\relay$ forwarding the received description without recovering the underlying information stream. In recover and forward, first the information is recovered at $\relay$ and then a proper description of it is forwarded to $\nodeB$. Finally, recompression is done by first recovering the information and then putting it together with the side information available at $\relay$ to compress it again. Note that availability of private and common (forwarded) messages at $\relay$ facilitate communication by providing side information for the $\nodeB$ to $\relay$ to $\nodeA$ communication.
  
  In order to accomplish the above, terminal $\nodeA$ uses a coding scheme similar to Wyner-Ziv by first covering $\X$ by $\auxU_0$, $\auxV_1$, $\auxU_1$ and $\auxW_1$ and then creating proper descriptions via binning. The four codebooks correspond to private message, simple forward, recover and forward and recompression schemes respectively. Therefore the binning for $(\auxU_0,\auxU_1,\auxW_1)$-triplet is done with respect to the side information at $\relay$, which results in the first term in (\ref{R1cascade}) and for $\auxV_1$, the binning is done with respect to $(\Y,\auxU_1)$. Upon receiving these bin indices, $\relay$ forwards the bin index corresponding to $\auxV_1$ to $\nodeB$, and then uses its side information to recover $\auxU_0$, $\auxU_1$ and $\auxW_1$. The relay keeps $\auxU_0$ as its own private message. To forward $\auxU_1$, $\relay$ creates a new description of it with respect to the side information at $\nodeB$ which can be done with any rate higher than $ I(\X;\auxU_1|\Y)$. Putting this together with the rate needed to forward the description of $\auxV_1$ to $\nodeB$, gives the first term in (\ref{R2cascade}). The recompression is done by treating $(\Z,\auxW_1)$ as a new super-source at the relay and using Wyner-Ziv coding by first covering it by $\auxS_1$ and then creating a proper description of it by binning it with respect to $(\Y,\auxU_1,\auxV_1)$ resulting in the second term in (\ref{R2cascade}). Upon receiving these, $\nodeB$ first uses its side information, $\Y$ to recover $\auxU_1$ and $\auxV_1$. It then uses $(\Y,\auxU_1,\auxV_1)$ to recover $\auxS_1$.
  
  Similarly the communication from $\nodeB$ to $\relay$ to $\nodeA$ is carried out by considering $\Y$ as the source, $(\auxU_1,\auxV_1,\auxS_1)$ as the transmitter's side information, $(\Z,\auxU_0,\auxU_1,\auxW_1,\auxS_1)$ as the side information at $\relay$ and $(\X,\auxU_0,\auxU_1,\auxV_1,\auxW_1)$ as the side information at $\nodeA$. Note that private message is only needed to improve the side information at $\relay$ for further communication from either terminal to $\relay$. Hence, there is no need to generate and send a private message. Using similar arguments, it is easy to show the inequalities (\ref{R3cascade}) and (\ref{R4cascade}).
\end{IEEEproof}

\section{Discussion}\label{discussion}
In this section, we introduce two scenarios to show that in general neither of the proposed schemes outperforms the other one. To argue this we consider some extreme cases. We assume all sources have discrete alphabets.

\subsection{Case 1: Large $\rate_{\nodeA\relay}$ and $\rate_{\nodeB\relay}$}
  Consider the case where $\rate_{\nodeA\relay}\geq H(\X|\Z)$ and $\rate_{\nodeB\relay}\geq H(\Y|\Z)$. This suggests that for both schemes, upon receiving $\msgM_{\indxt\relay}$, $\indxt=\nodeA,\nodeB$, the relay can reconstruct the respective source losslessly. With DSC/delivery scheme it is easy to see that $\rate_{\relay\nodeA}=\rate_{\Y|\X}(\D_\nodeA)$ and $\rate_{\relay\nodeB}=\rate_{\X|\Y}(\D_\nodeB)$ are achievable. Note that these rates are the same as the ones in Remark \ref{DSC delivery - simple outer}, which proves the inner bound is tight in this case. Comparing these rates with the outer bound for two cascaded rounds in Theorem \ref{cascade outer bound} we observe that they fall outside or in the worst case on the outer bound for two cascaded rounds scheme. This means that DSC/delivery can achieve lower rates than two cascaded rounds.

\subsection{Case 2: Large $\rate_{\relay\nodeA}$ and $\rate_{\relay\nodeB}$}
  This time let $\rate_{\relay\nodeA}\geq H(\Y,\Z|\X)$ and $\rate_{\relay\nodeB}\geq H(\X,\Z|\Y)$. This allows for the relay to forward any description of the sources that it receives as well as its own side information $\Z$ losslessly to each terminal, in both communication schemes. In this case two cascade rounds scheme becomes equivalent to interactive source coding with multiple rounds \cite{Kaspi1985}. Therefore
\begin{IEEEeqnarray*}{l}
  \rate_{\nodeA\relay}\geq I(\X;\auxU_1|\Y,\Z)\\
  \rate_{\nodeB\relay}\geq I(\Y;\auxU_2|\X,\Z,\auxU_1)
\end{IEEEeqnarray*}
for some $\auxU_1$ and $\auxU_2$ satisfying $\auxU_1-\X-(\Y,\Z)$ and $\auxU_2-(\Y,\Z,\auxU_1)-\X$ and some decoding functions $\hatX=\decod_\nodeB(\Y,\Z,\auxU_1)$ and $\hatY=\decod_\nodeA(\X,\Z,\auxU_1,\auxU_2)$ such that $E\fd_\nodeA(\Y,\hatY)\leq\D_\nodeA$ and $E\fd_\nodeB(\X,\hatX)\leq\D_\nodeB$ is achievable. Note that this inner bound may not be tight in this case. Nevertheless, setting $\auxU_2$ independent of $\auxU_1$ and $\auxU_2-\Y-(\X,\Z)$ and optimizing over $\auxU_1$ and $\auxU_2$ we can achieve the outer bound for DSC/delivery in Remark \ref{DSC delivery - simple outer}. Therefore the inner bound for two cascaded rounds can potentially achieve lower rates than DSC/delivery for some sources and distortion functions.


\section{Conclusion}

In this paper we have studied different possible schemes, DSC/delivery and two cascaded rounds, for interactive relay assisted source coding, and characterized inner and outer bounds for each scheme. We have evaluated these bounds in some extreme cases to show that in general neither of the schemes dominates the other. Future work includes identifying scenarios in which the inner and outer bounds are tight.



\bibliographystyle{IEEEtran}
\bibliography{GlobalSIPref}


\end{document}